\documentclass{article}
\newcommand{\lisn}{A\&O, LISN, Université Paris-Saclay, CNRS, Inria TAU, France}
\newcommand{\ucsd}{SCCN, INC, SDSC, University of California San Diego, USA}
\newcommand{\rad}{Donders Institute, Radboud University, Netherlands}
\newcommand{\google}{Google DeepMind, ChaLearn, USA}
\newcommand{\child}{Child Mind/Nathan Kline Institutes, New York, USA}
\newcommand{\ben}{Ben-Gurion University of the Negev, Israel}
\newcommand{\meta}{FAIR Brain \& AI team/Reality Labs, Meta, France}

\newcommand{\ufabc}{Federal University of ABC, Brazil}
\newcommand{\nerv}{Inria NERV, Paris Brain Institute, France}
\newcommand{\cneuro}{Cuban Neuroscience Center, Cuba and China}
\newcommand{\mcgill}{McGill University, Canada}
\newcommand{\cnrs}{CNRS, France}

\PassOptionsToPackage{numbers, compress}{natbib}

\usepackage[preprint]{neurips_2025}

\usepackage[multiple]{footmisc} 
\usepackage[utf8]{inputenc} 
\usepackage[T1]{fontenc}    

\usepackage{hyperref}
\usepackage{xcolor}
\usepackage{url}            
\usepackage{booktabs}       
\usepackage{amsfonts}       
\usepackage{nicefrac}       
\usepackage{microtype}      
\usepackage{xcolor}         
\usepackage{ulem}
\usepackage{comment}
\usepackage{booktabs}
\usepackage{array}
\usepackage{tabularx}
\usepackage{multirow}
\usepackage{amsmath} 
\usepackage{amsfonts}       
\usepackage{xspace}
\usepackage{todonotes}
\usepackage{xparse}
\usepackage{multicol}
\usepackage{enumitem}
\usepackage{ragged2e}  
\usepackage{caption}
\justifying

\definecolor{myblue}{HTML}{0064E0}
\definecolor{trb}{RGB}{114,142,187} 
\definecolor{factp}{RGB}{129,100,149} 
\definecolor{lbg}{RGB}{141,177,111} 

\hypersetup{
    colorlinks=true,
    linkcolor=myblue,     
    citecolor=myblue,     
    urlcolor=myblue       
}
\makeatletter
\long\def\@makefntext#1{%
  \noindent
  \makebox[1.8em][r]{\@makefnmark\,}%
  \justifying 
  #1%
}
\makeatother
\newcolumntype{L}[1]{>{\raggedright\arraybackslash}p{#1}}

\makeatletter
\renewcommand{\@makefntext}[1]{%
  \noindent
  \makebox[0em][l]{\@thefnmark}#1}
\makeatother

\newcommand{\tr}{\ensuremath{\mathrm{train}}}
\newcommand{\test}{\ensuremath{\mathrm{test}}}
\newcommand{\va}{\ensuremath{\mathrm{valid}}}
\newcommand{\X}{\ensuremath{\mathcal{X}}\xspace}
\newcommand{\Y}{\ensuremath{\mathcal{Y}}\xspace}

\newcommand{\Ph}{\ensuremath{\mathcal{P}}\xspace}
\newcommand{\Test}{\ensuremath{\mathrm{Testing}}}
\newcommand{\Valid}{\ensuremath{\mathrm{Validation}}}
\newcommand{\Train}{\ensuremath{\mathrm{Training}}}

\newcommand{\y}{\ensuremath{\mathbf{Y}}}

\newcommand{\channel}{\ensuremath{c}}
\newcommand{\timestep}{\ensuremath{t}}
\newcommand{\event}{\ensuremath{k}}
\newcommand{\factor}{\ensuremath{\mathbf{p}_{fac}}}

\newcommand{\metric}{\ensuremath{\mathcal{S}}}

\newcommand{\rmse}[1]{\ensuremath{\text{RMSE}\left(#1\right)}}
\newcommand{\nrmse}[1]{\ensuremath{\text{nRMSE}\left(#1\right)}}
\newcommand{\std}[1]{\ensuremath{\text{std}\left(#1\right)}}

\newcommand{\true}{\ensuremath{\text{true}}}
\newcommand{\pred}{\ensuremath{\text{pred}}}

\newcommand{\x}{\ensuremath{\mathbf{X}}}

\makeatletter

\def\@fnsymbol#1{%
  \ensuremath{%
    \ifcase#1%
      \or *
      \or \dagger
      \or \ddagger
      \or \mathsection
      \or \mathparagraph
      \or \|
      \or \clubsuit
      \or \maltese
      \or \lozenge
      \or 
      \or \spadesuit
      \or \heartsuit
      \or \diamondsuit
      \or \ast\ast
      \or $\dagger\dagger\dagger$
    \else
      \@ctrerr%
    \fi%
  }%
}
\makeatother

\makeatletter
\newcommand{\getfnsymbol}[1]{\@fnsymbol{#1}}
\makeatother

\title{EEG Foundation Challenge: \\ From Cross-Task to Cross-Subject EEG Decoding}

\author{
\textbf{Bruno Aristimunha}\footnotemark[1]\, \footnotemark[2]\,\,\footnotemark[11]\and
\textbf{Dung Truong}\footnotemark[3]\and
\textbf{Pierre Guetschel}\footnotemark[4]\and
\textbf{Seyed Yahya Shirazi}\footnotemark[3]\and
\textbf{Isabelle Guyon}\footnotemark[5]\and
\textbf{Alexandre R. Franco}\footnotemark[6]\and
\textbf{Michael P. Milham}\footnotemark[6]\and
\textbf{Aviv Dotan}\footnotemark[7]\and
\textbf{Scott Makeig}\footnotemark[3]\and
\textbf{Alexandre Gramfort}\footnotemark[8]\and
\textbf{Jean-Remi King}\footnotemark[8]\and
\textbf{Marie-Constance Corsi}\footnotemark[2]\and
\textbf{Pedro A. Valdés-Sosa}\footnotemark[12]\and
\textbf{Amit Majumdar}\footnotemark[3]\and
\textbf{Alan Evans}\footnotemark[13]\and 
\textbf{Terrence J Sejnowski}\footnotemark[3] \and 
\textbf{Oren Shriki}\footnotemark[7] \and 
\textbf{Sylvain Chevallier}\footnotemark[1]\and
\textbf{Arnaud Delorme}\footnotemark[3]\, \footnotemark[9] \\
{\centering\tt\href{mailto:neurips2025-eeg-competition@googlegroups.com}{neurips2025-eeg-competition@googlegroups.com} \vspace{-0.5em}}
}

\begin{document}

\maketitle

\footnotetext{\textsuperscript{\getfnsymbol{1}}\lisn
        \hspace{0.4em}  \textsuperscript{\getfnsymbol{2}}\nerv}
\footnotetext{\textsuperscript{\getfnsymbol{3}}\ucsd \hspace{1.2em} \textsuperscript{\getfnsymbol{9}}\cnrs
        }
\footnotetext{\textsuperscript{\getfnsymbol{4}}\rad
    \hspace{5.25em}
 \textsuperscript{\getfnsymbol{5}}\google}
\footnotetext{\textsuperscript{\getfnsymbol{6}}\child
       \hspace{5.3em} \textsuperscript{\getfnsymbol{7}}\ben} 
\footnotetext{\textsuperscript{\getfnsymbol{8}}\meta
       \hspace{5.4em} \textsuperscript{\getfnsymbol{11}}\ufabc}
\footnotetext{\textsuperscript{\getfnsymbol{12}}\cneuro 
        \hspace{7.65em} \textsuperscript{\getfnsymbol{13}}\mcgill}

\begin{abstract}
Current electroencephalogram (EEG) decoding models are typically trained on small numbers of subjects performing a single task.
Here, we introduce a large-scale, code-submission-based competition comprising two challenges.
First, the Transfer Challenge asks participants to build and test a model that can zero-shot decode new tasks and new subjects from their EEG data. 
Second, The Psychopathology factor prediction Challenge asks participants to infer subject measures of mental health from EEG data. 
For this, we use an unprecedented, multi-terabyte dataset of high-density EEG signals (128 channels) recorded from over 3,000 child to young adult subjects engaged in multiple active and passive tasks. 
We provide several tunable neural network baselines for each of these two challenges, including a simple network and demographic-based regression models. 
Developing models that generalize across tasks and individuals will pave the way for ML network architectures capable of adapting to EEG data collected from diverse tasks and individuals. Similarly, predicting mental health-relevant personality trait values from EEG might identify objective biomarkers useful for clinical diagnosis and design of personalized treatment for psychological conditions.
Ultimately, the advances spurred by this challenge could contribute to the development of computational psychiatry and useful neurotechnology, and contribute to breakthroughs in both fundamental neuroscience and applied clinical research.
\end{abstract}


\section{Competition description}

\begin{figure}[!ht]
\scriptsize
    \centering
\includegraphics[width=\linewidth]{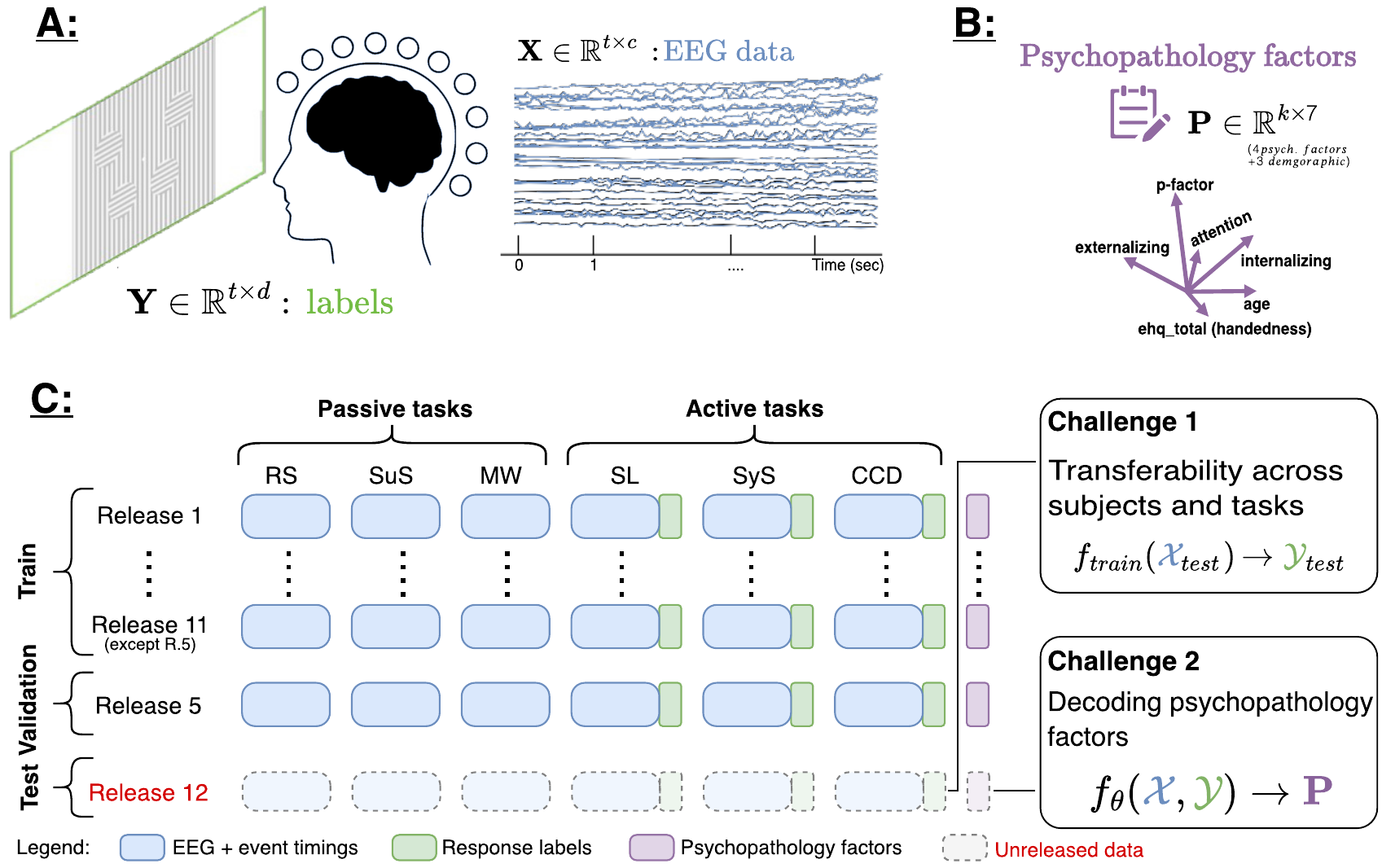}
    \caption{HBN-EEG Dataset and Data split. \textbf{A.} EEG is recorded using a 128-channel system during active tasks (i.e., with user input) or passive tasks. \textbf{B.} The psychopathology and demographic factors. \textbf{C.} The datasets split into Train, Test and Validation. Details in \autoref{sec:data}.}
    \label{fig:workflow}
\end{figure}

We propose the \textbf{EEG Foundation Challenge} to advance research on developing \emph{generalized} and \emph{transferable} neural representations for EEG decoding (\autoref{fig:workflow}). This initiative focuses on identifying mental biomarkers and decoding cognitive tasks across \emph{large} number of subjects, addressing the fundamental challenge of interpreting complex brain signals associated with key psychopathological factors and experimental stimuli.

In recent years, the neuroscience community has increasingly looked toward Large Language-Model-style architectures to advance brain activity decoding \cite{zhou2025brain}. 
However, this paradigm shift remains limited by the data scale and structure of currently available data \cite{bomatter2025limited, Gemein2025}. Unlike the massive, richly annotated corpora used to train language models, EEG datasets are relatively small and lack comparable hierarchical organization. 
Moreover, brain signal data present intrinsic challenges, including differences in electrode montages, recording protocols, various methods for splitting the data, and underlying neurological factors \cite{aristimunha2024best, nerve, delorme2023eeg, kessler2024eeg, trubutschek2024eegmanypipelines}.  
These challenges have hindered the development of broad, generalizable, and cross-task models that extend beyond simple classification, highlighting the need for larger, more integrated EEG datasets to support next-generation research.

A given EEG dataset usually corresponds to one specific cognitive task or condition with only a single type of ground-truth label, often requiring a separate model to be trained for each task. 
In practice, models are frequently developed and tuned in isolation on each dataset—and sometimes even on each individual subject—due to pronounced physiological differences between subjects \citep{Apicella2022-kv,Mikkelsen2021-zy,Saha2019-zz}. 
This conventional single-task, -subject training and evaluation fails to leverage commonalities across datasets and yields models that struggle to generalize beyond the context in which they were trained.

For these reasons, we argue that  \emph{a general-purpose EEG foundation competition is needed to move beyond the one-task-per-model paradigm}. 
A unified model could simultaneously decode both enduring mental traits and dynamic cognitive states, bridging traditionally separate objectives. 
For instance, it could infer a participant’s latent mental health score while also predicting task performance from the same EEG stream. 
Training on diverse tasks and subjects would enable the model to learn robust representations that generalize across users and adapt to different electrode configurations. 
As model capacity and data diversity scale, we expect \emph{more reliable regression of cognitive and clinical biomarkers}, even in the presence of noise.

Previous EEG decoding challenges have shown that well-designed competitions can drive progress in brain decoding. Our BEETL Motor Imagery Challenge at NeurIPS 2021 \citep{beetl2022} gathered 130+ contestants from 40 research groups and over 1,382 submissions. Our Brain Age Prediction Challenge 2022 \citep{age2022} attracted 200+ participants from 40 countries, generating 30–100 daily submissions. In 2023, our Sleep States Competition \citep{sleep2023} engaged 80 teams from 20 countries. In 2024, we co-organized a large-scale challenge with the Child Mind Institute \citep{child-mind-institute-problematic-internet-use}, focusing on predicting problematic internet use in youth using accelerometry, assessments, and questionnaires. It drew 12,912 registrants, 2,436 active contestants across 1,877 teams, and 38,329 submissions. This year, we are currently running additional challenges on the Healthy Brain Network's fMRI and actigraphy. The ongoing WiDS Datathon 2025 \citep{widsdatathon2025}, part of the Women in Data Science initiative, has already reached 5,062 contestants, with 1,614 active contestants across 703 teams and 9,470 submissions to date. These results highlight our strong track record of organizing impactful, large-scale challenges and mobilizing the machine learning research community for an appealing NeurIPS competition.

\subsection{Novelty}
The EEG Foundation Challenge is inspired by and builds up on our NeurIPS 2021 BEETL challenge, with a bigger scope and ambition. Unlike BEETL, which focuses on specific tasks (sleep staging and motor imagery), the present competition introduces \emph{an unprecedented combination of scale, complexity, and objectives}. The competition features three major innovations:

1) \textbf{Unprecedented Scale and Complexity:} The competition's dataset encompasses over 3,000 participants with high-density 128-channel EEG recordings – an order of magnitude larger than typical EEG challenge datasets. Each participant engaged in six distinct cognitive tasks (ranging from resting-state to active learning and attention tasks), providing a rich multi-task, multi-condition collection of neural data. This participant, metadata, and recording breadth and diversity far \emph{exceeds that of any prior EEG competition} \citep{sleep2023, beetl2022, age2022}. So, models should handle heterogeneous time-series data conditions, learn generalizable representations across tasks and participants. EEG decoding research has progressed beyond traditional discrete classification methods, with novel approaches now addressing more complex representational challenges at this large scale \cite{zhou2025brain, bomatter2025limited, banville2025scaling, wang2025lead, turgut2025foundation, kiessner2024reaching, d2024decoding, gemein2020machine}.

2) \textbf{Zero-Shot Cross-Domain Generalization:} The cross-task transfer learning scenario in EEG decoding is remarkably underexplored \cite{aristimunha2023evaluating}. Our Challenge 1 addresses a key goal in neurotechnology: decoding cognitive function from EEG without explicit behavioural labels. Participants must develop \emph{models capable of zero-shot generalization across both novel tasks and novel subjects}. The competition is structured to reward models that learn domain-invariant and subject-invariant representations. This means a submitted model might be trained on a subset of tasks and then tested on data from a held-out task or condition, evaluating its capacity to generalize without task-specific fine-tuning. Notably, no existing benchmarks or datasets have been developed to address these issues \cite{chevallier2024largest, ding2021learning}.

3) \textbf{Prediction of Latent Psychological Constructs:} In contrast to past challenges, our Challenge 2 targets latent psychological constructs as prediction outputs. Specifically, participants will predict transdiagnostic psychopathology factors (such as the p-factor) derived from standardized clinical questionnaires \citep{Scopel-Hoffmann2022-sy,Caspi2014-yf}. While the p-factor is intended to reflect a psychological trait \cite{Caspi2014-yf}, EEG could also be modulated by fluctuating  ``states'' such as engagement, fatigue, and other time-variant components \cite{bomatter2024machine, sabbagh2020predictive}. These components may be implicitly encoded in neural activity, helping to account for such variabilities in the data. The challenge holds considerable potential for scientific impact, particularly if participating teams could improve decoding of psychopathology factors from EEG data, thereby offering empirical support for EEG-based biomarkers for mental health \cite{hong2023general}.

By combining unexplored targets, fundamental scientific questions, and a large-scale, high-dimensional dataset, this competition is a unique opportunity to advance the state of EEG-based predictive modelling and to use machine learning in the discovery of psychiatric biomarkers. Beyond its methodological and scientific contributions, this challenge has meaningful societal implications: identification of psychopathology metrics through accessible, non-invasive measures like EEG could \emph{contribute to earlier detection, more precise monitoring, and ultimately more effective interventions for mental health conditions}. As mental health continues to emerge as a global public health priority, innovations in scalable and data-driven assessment methods can inform clinical practice and policy, reduce stigma, and improve outcomes for individuals affected by psychiatric disorders.

\subsection{Data}\label{sec:data}\vspace{-0.5em}





The competition leverages the Healthy Brain Network Electroencephalography (HBN-EEG) datasets~\cite{Shirazi2024-ye}. 
It is made of 12 dataset releases in total, grouped according to the time of release, 11 releases shared, and one withheld and unreleased for competitions. This large-scale collection contains high-density (128-channel) EEG recordings from children and young adults aged 5-21 years with more details described at \citep{Shirazi2024-ye}. 
The dataset is formatted according to the Brain Imaging Data Structure (BIDS) standard \citep{Gorgolewski2016-vj, Pernet2019-vh} and includes comprehensive event annotations using Hierarchical Event Descriptors (HED) \citep{Robbins2021-rf, makeig2024events}, making it particularly suitable for cross-task analysis and machine learning applications. 

Formally, let $X \in \mathbb{R}^{\channel \times \timestep}$ be an EEG recording segment (\autoref{fig:workflow}), with $\channel=128$ the number of electrodes and $\timestep$ the time steps, 
$Y \in \mathbb{R}^{\event \times \timestep}$ be the associated labels, with $\event$ the number of event labels, and $P \in \mathbb{R}^7$ be subject's traits including 3 demographic attributes and 4 psychological factors, e.g., p-factor \factor. 
A dataset is defined as $\left( \textcolor{trb}{\X}, \textcolor{lbg}{\Y}, \textcolor{factp}{\Ph} \right)$, with $N$ EEG recording from $S$ subjects, $\textcolor{trb}{\X}=\left\{ \textcolor{trb}{\x}_i \right\}_N$ with their associated labels $\textcolor{lbg}{\Y}=\left\{ \textcolor{lbg}{Y}_i \right\}_N$ and psychological factors and demographic attributes $\textcolor{factp}{\Ph}=\left\{\textcolor{factp}{\mathbf{p}} \right\}_7$ for each $s$ subject.

We consider here different datasets with different types of labels.
In this competition, the trial timing is assumed to be known, targeting offline applications that are common in the medical context, leaving real-time decoders outside the scope of this competition. We assume that the dataset is randomly split into a training set of size $N_\tr$, a validation set of size $N_\va$ and a test set of size $N_\test$.
For training and validation sets, both \textcolor{trb}{\X} and \textcolor{lbg}{\Y}  are available, while for test sets, only \textcolor{trb}{\X} is available to the contestant.
The general decoding problem amounts to using the training set $\textcolor{trb}{\X}_\tr$ to learn a model $f_\theta$ with parameters $\theta$ mapping each trial $\textcolor{trb}{\x}$ to the associated label $\textcolor{lbg}{\y}$.

\subsubsection{Dataset Description}

The competition data set includes EEG recordings from more than $3,000$ participants in six distinct cognitive tasks, divided into passive -- Resting State (RS), Surround Suppression (SuS), Movie Watching (MW) -- and active -- Contrast Change Detection (CCD), Sequence Learning (SL), Symbol Search (SyS) --  categories. 
SuS and CDD rely on stimulus that are similar to those employed in SSVEP tasks.

The data of each participant are accompanied by four dimensions of psychopathology (internalization, externalization, attention, and p-factor) derived from a bifactor model of parent-reported questionnaire responses to the Child Behavior Checklist (CBCL) \citep{Achenbach1999-ey,Scopel-Hoffmann2022-sy}. 
These psychopathology factors represent orthogonal dimensions of mental health and serve as target variables for the regression component for \textbf{Challenge 2} \cite{Scopel-Hoffmann2022-sy}.
Additionally, the age, gender, and handedness (scored using the Edinburg Inventory, EHQ-Total) \citep{Oldfield1971-dq} are included as part of the demographic information for each participant. 
While some psychopathology factors, like internalizing and externalizing, are influenced by a participant’s age, other factors such as the p-factor and attention, are largely unaffected by age \autoref{fig:workflow}B. \vspace{-0.5em}

\paragraph{Data Access and Ethics}
The first 11 releases of the HBN-EEG dataset are freely available under CC-BY-SA-4.0 licence through \href{https://nemar.org}{NEMAR}~\citep{Delorme2022-sl}. $\Test$ set use release 12 that is currently \textbf{unreleased}. The Chesapeake Institutional Review Board approved the data collection, with written informed consent obtained from participants 18 years or older and from legal guardians of younger participants. All participants were anonymized using the NIMH's Global Unique Identifier Tool (GUID) \citep{Johnson2010-pu} to ensure no exposure to personally identifiable information.

\subsubsection{Competition Data Split} \label{sec:datasplit}

For this competition, we provide contestants with a $\Train$ set, complete EEG recordings and psychopathology measures from approximately $3,000$ subjects, releases 1-11, except for the Release 5. The $\Valid$ set, Release 5 comprises data from the remaining subjects for the teams to test their models. 
This split ensures sufficient data for model training while maintaining a robust evaluation set for assessing cross-task generalization and psychopathology prediction performance. The final evaluation will be performed using the withheld test set, the Release 12 that is not available publicly, but will be released publicly in the foreseeable future after the competition.

As contestants submit their code, they will not have direct access to the test set. No subject's data is spread across multiple releases; this means the $\Train$, $\Valid$, and $\Test$ are completely separate in terms of subjects, ensuring no feature leakage in this competition, avoiding subjects' co-founders.
An illustration of the data split can be found in \autoref{fig:workflow}.

For the Challenges, participants are strongly encouraged to leverage additional datasets beyond the provided training set, particularly those involving SSVEP stimuli. As a starting point, we recommend exploring the Mother of All BCI Benchmarks (MOABB) library \cite{chevallier2024largest, Aristimunha_Mother_of_all_2023}, which provides a wide range of small, but well-curated EEG datasets. However, contestants are responsible for designing models capable of adapting to variations in channel configurations and signal characteristics across datasets.


\subsection{Tasks and application scenarios}\label{task-and-application}



The competition focuses on two primary challenges that address significant problems in neurophysiological research and clinical neuroscience: \emph{inferring mental states} and \emph{inferring mental traits}  from EEG signals.
From a machine learning perspective, the model needs to generalize across two challenges: a \emph{transfer learning} problem and a \emph{regression} problem.
In \textbf{Challenge 1}, the source domain is the concatenation of multiple datasets with multiple tasks on train subjects, and the target domain is a specific task on test subjects, whereas in \textbf{Challenge 2}, the objective is to predict a specific psychological trait from EEG recording without labels and with the other subject's traits: 
\vspace{0.1em}
\begin{gather*}
    \text{Challenge 1 - } f_\theta: \left( \textcolor{trb}{\X}, \textcolor{factp}{\Ph} \right) \to \textcolor{lbg}{\Y} , \;\;\; 
    \text{Challenge 2 - } f_\theta: \left( \textcolor{trb}{\X}, \textcolor{lbg}{\Y} \right) \to \ \textcolor{factp}{\Ph},
\end{gather*}
\vspace{-1.5em}


In \textbf{Challenge 1 - Transferability across subjects and cognitive tasks}, the contestants are provided with multiple databases with subjects performing similar tasks, namely SSVEP with ERP components, to train their model on the HBN Releases datasets. For this challenge, the contestants must predict behavioural performance metrics in an active task (Contrast Change Detection, CCD): response time relative to the start of contrast change index on the stimulus time with a 500 ms shift, i.e., 
The window period is 0.5 to 2.5 seconds after stimulus onset for prediction.

At the inference time, we will use the Contrast Change Detection data with the two-second data epochs and plus their demographic at the epoch level, as indicated in \autoref{fig:workflow}.

Detailed data split, epoching, and basic scripts for data loading and curation are available at the competition website.
Contestants should submit per-trial predictions. The metrics used to evaluate the predictions are detailed in Sect~\ref{sec:metrics1}.

This challenge has significant implications for clinical and research settings as obtaining and retaining EEG recordings and accounting for participant's response variability may be difficult or impossible due to participants' or environment limitations \citep{DiStefano2019-cq,Demazure2021-jf}.
The experiment length and actively engaging participants in experiments are challenging for several populations, including children, individuals with severe motor impairments, or those with communication and developmental disorders \citep{DiStefano2019-cq, Zech2022-sw, Reilly2019-ly}.
Passive EEG paradigms such as movie watching with shorter duration that can predict performance capabilities would enable more inclusive and accessible neurological assessments. 
Additionally, by extracting maximal information from minimal recording sessions, researchers can reduce the time and effort required from participants, making neurophysiological research more accessible and efficient.

In \textbf{Challenge 2 - Decoding psychopathology factor from the EEG}\label{sec:challenge2}, the contestants will have to regress the continuous four psychopathology scores 
using EEG recordings from multiple cognitive tasks. 
For this challenge, only subjects with at least 15 minutes of total EEG data and complete demographic information will be included for validation and test (>78\% of the participants).
Contestants should predict per-subject score. 
The metrics used to evaluate the predictions are detailed in \autoref{sec:metrics2}.

This challenge represents a novel approach to mental health assessment that leverages objective neurophysiological measures instead of relying on subjective self-reports or clinical observations. 
It addresses critical challenges in mental health assessment and treatment. 
Current psychiatric diagnoses mainly rely on subjective assessments, leading to potential inconsistencies and biases~\cite{loo2016research}. 
EEG-based biomarkers could provide complementary, physiologically-grounded measures of mental health  \cite{bomatter2024machine}. 
Neurophysiological signatures of psychopathology may be detectable before clinical symptoms fully manifest, potentially enabling earlier intervention.
By predicting continuous psychopathology dimensions rather than categorical diagnoses, this approach aligns with modern transdiagnostic frameworks like the Research Domain Criteria that conceptualizes mental health along continuous dimensions \cite{Insel2010-fa,Cuthbert2015-ks}.
Also, objective EEG-based measures could be used as biomarkers to track treatment response over time, providing quantitative feedback on intervention efficacy.

\textbf{Scientific and Technical Challenges:}
Both challenges raise substantial scientific and technical questions that are central to the development of EEG foundation models. EEG signals are inherently high-dimensional, noisy, and temporally complex, with considerable inter-subject and inter-task variability. 
Building models that generalize across subjects, tasks, and recording (electrode) configurations requires scalable architectures capable of robust pattern recognition in the face of such variability.

\subsubsection{Ethical Considerations}

To consider the potential ethical implications of these challenges, we implemented the following steps: \textbf{1)} We ensured that all data have collected with appropriate consent. \textbf{2)} We focus here on psychological traits instead of diagnostic categories, to mitigate the risks of diagnostic misuse and to limit the potential for discriminatory use of EEG-based predictions. \textbf{3)} We aim to develop models that perform equitably across different populations by using a diverse dataset and including various demographic factors.

The development of highly accurate, task- and subject-general models has the potential to improve the identification of interpretable neural biomarkers, thereby supporting clinical insight and decision-making by health professionals.
The competition challenges align with the NeurIPS community's interests in \emph{developing robust, generalizable machine learning methods for complex, high-dimensional time series data while addressing significant real-world challenges in healthcare and cognitive science}. \vspace{-0.5em}

\subsection{Metrics}\vspace{-0.5em}

We employ carefully selected metrics to evaluate performance on each challenge, ensuring they reflect real-world utility and statistical robustness. \vspace{-0.5em}

\paragraph{Challenge 1}\label{sec:metrics1}
Contestants will predict behavioural performance, i.e., the \textbf{response time}, in the Contrast Change Detection (CCD) task using EEG data from the Surround Suppression (SuS) task and pre-trial EEG. 
We decided not to use correct/incorrect classifications for this challenge.
The evaluation metric for this regression challenge is the normalized root mean square error $\nrmse{\cdot, \cdot}$ between the true and predicted response time.
The normalized root mean square error is the metric quantifying prediction error magnitude, providing a clinically interpretable measure of prediction deviation
Formally, this metric is defined as follows:
\begin{align}
    \rmse{\y_\true, \y_\pred} &= \sqrt{\frac{1}{N}\sum_{i=1}^N\left| \y^i_\true-\y^i_\pred\right|^2 } \\
    \std{\y} &= \sqrt{\frac{1}{N}\sum_{i=1}^N\left| \y^i-\overline{\y}\right|^2 }\\
    \nrmse{\y_\true, \y_\pred} &= \frac{\rmse{\y_\true, \y_\pred}}{\std{\y_\true}} \\
    \metric_1 &= \nrmse{\y_{1,\true}, \y_{1,\pred}}
    \label{eq:metric1}
\end{align}
With $\y_{1,\true}$ being the vector containing the true response times, $\y_{1,\pred}$ the predicted ones, and $\metric_1$ the score for challenge 1.


\paragraph{Challenge 2}\label{sec:metrics2}
Contestants will develop models to predict the psychopathology factors from EEG recordings. 
The evaluation metric for this challenge is also the normalized root mean square error:
\begin{equation}
    \metric_2 = \nrmse{\y_{2,\true}, \y_{2,\pred}}
\end{equation} 
With $\y_{2,\true}$ being the labels vector containing the p-factors obtained from questionaries, $\y_{2,\pred}$ containing the p-factors predicted from the EEG recordings, and $\metric_2$ the score for challenge 2.

\paragraph{Overall Ranking}

The objective of this competition being to foster the development of foundation models, the overall ranking of this competition is a combination of both challenges to promote the design of models able to address multiple tasks. The final score $ \metric_{\text{overall}}$  
reflects the greater clinical significance of psychopathology prediction. This comprehensive evaluation framework ensures that winning approaches demonstrate robust performance across multiple clinically relevant dimensions, encouraging the development of models with real-world utility for neurophysiological assessment and mental health applications.
\begin{equation}
    \metric_{\text{overall}} = 0.3 \, \metric_1 + 0.7 \, \metric_2
\end{equation}

\subsection{Baselines, code, and material provided}


The starter kit can be found at: \href{https://eeg2025.github.io/baseline/}{https://eeg2025.github.io/baseline/}. This code relies on the 
\href{https://braindecode.org}{\textsc{Braindecode}} and \href{https://eegdash.org}{\textsc{EEGDash}} libraries. These libraries allow data search, loading, fetching, and readily applying deep learning methods on EEG data. 
In particular, \textsc{Braindecode} bridges the \textsc{MNE-Python} \cite{gramfort2014mne} and PyTorch \cite{torch} libraries (respectively made for brain signal processing and deep learning). 
\textsc{EEGDash} allows to train \textsc{Braindecode} models or any \textsc{PyTorch} model from data retrieved from \href{https://openneuro.org}{\textsc{OpenNeuro}} and \href{www.nemar.org}{\textsc{NEMAR}}. In the \textsc{Braindecode} library, we provide a \textsc{PyTorch} model zoo for EEG decoding featuring over 30 deep learning models—including ShallowConvNet \cite{schirrmeister2017deep}, EEGNet \cite{lawhern2018eegnet}, EEG-Inception \cite{santamaria2020eeg}, EEG-Conformer \cite{song2022eeg}, EEGNex \cite{CHEN2024105475}, ATCNet \cite{altaheri2022physics}, BIOT \cite{yang2023biot}, Labram \cite{jiang2024large} and more, all rigorously standardized and tested for correctness. The provided scripts are only for reference, and the contestants can freely use their codebase for development.


\subsection{Website, tutorial and documentation}\label{sec:website}




All information regarding this competition will be provided at \url{https://eeg2025.github.io}. This website will contain a description of the data, links and command lines to download the data, a script allowing the load the data and run the baseline(s), the timeline, the leaderboard,
FAQ, and news. 
We also provide a dedicated \href{mailto:neurips2025-eeg-competition@googlegroups.com}{e-mail} address for communication with contestants. We will create a GitHub Community Forum for the competition to facilitate communication between participants and organizers, as well as among participants themselves. 

\section{Organizational aspects}

\subsection{Protocol}



Contestants are provided will all necessary scripts to download the data. Some model examples are provided, but they could choose freely their model. They need to train their model on their own computational infrastructure, we provide example how to use Google Colab for contestant without specific resources. 
They need to create a team account on the \href{https://www.codabench.org/}{\textsc{Codabench}} platform and to provide a code submission with their trained model. 
The Université Paris-Saclay and San Diego Supercomputer Center will provide the compute worker nodes to run the model against test.

The competition will have two phases: in the \textbf{warm-up phase}, the contestants start to submit their code, the evaluation will be performed on \emph{validation set} (HBN Release 5) that is publicly available, allowing contestants to verify that their code is running correctly. No limit will be fixed to the number of submissions to facilitate debugging. 
This phase should serve as an incentive to incite contestants to join the competition, with an easy to beat baseline and several available models, ready to be tuned.

During the second \textbf{final phase}, the contestants' model will be evaluated on \emph{test set} (HBN Release 12, not publicly available). The number of submissions per day will be limited to avoid overfitting the leaderboard. Their final rankings and scores will be released after the competition deadline. Each team will upload a 2-page document, in paper format with methods, analysis and discussion.
\vspace{-0.5em}

\subsection{Rules and Engagement}






The rules that will be provided to the contestants are the following: \vspace{-0.5em}

\begin{itemize}[leftmargin=1em,itemsep=0ex]
    \item Contestants are \emph{allowed and encouraged to use any datasets} to pre-train.     
    \item Contestant submit their code during the inference stage, this is a \emph{code submission competition}.
    \item \emph{Related members} from the organising team can participate, but \emph{are ineligible for prizes}.
    \item The top 10th teams will have their code release after the final submission.
\end{itemize}\vspace{-1.0em}

\subsection{Schedule and readiness}
\vspace{-0.5em}
The tentative timeline for challenge preparation is:\vspace{-0.5em}

\begin{itemize}[leftmargin=1em,itemsep=0ex]
    \item \textbf{[done]} Finishing dataset interface with \textsc{OpenNeuro} and \textsc{NEMAR} portals. 
    \item \textbf{[done]} Finishing baseline integration using \textsc{Braindecode} models.
    \item \textbf{[done]} Verifying the reproducibility of the scripts. 
    \item \textbf{30/04/2025}: Cleaning up the script for releasing.
    \item \textbf{17/05/2025}: Extend CodaBench configuration to run on NSF's ACCESS or bigger cluster.
    \item \textbf{18/05/2025}: Complete internal beta testing to identify and address potential bottlenecks.
    \item \textbf{18/05/2025}: Announcement of the challenge on social media and our dissemination channels as describe at \autoref{sec:promotion}.
\end{itemize}

The proposed timeline for the challenge is as follows: 
\begin{itemize}[leftmargin=1em,itemsep=0ex]
    \item \textbf{15/07/2025}: 
    \textbf{Warm-up Phase} started.
    \item \textbf{15/09/2025}: \textbf{Final Phase} started with unreleased data.
    \item \textbf{31/10/2025}: End of the competition.
    \item \textbf{30/11/2025}: Competition reports and competition analysis paper. 
    \item \textbf{At NeurIPS}: Organization of the competition session, with keynote about the lessons learned. 
\end{itemize}\vspace{-1em}

\subsection{Competition promotion and incentives}\label{sec:promotion}



We anticipate participation from three key groups: \textbf{(1)} researchers with proven success in EEG decoding, particularly transfer learning or psychopathology detection; \textbf{(2)} researchers seeking to benchmark strong, unpublished approaches on a high-visibility platform like NeurIPS; and \textbf{(3)} machine learning experts, especially from transfer learning domains, interested in applying their methods to EEG data.

To reach these communities, we will directly contact over 4,000 participants from our five previous decoding competitions and disseminate the call via major mailing lists (including EEGLAB with 17,000+ members, MNE-Python, and NeuroTechX). We will further leverage the organizing team's extensive professional networks across academia and industry for broad promotion. To incentivize strong participation, significant awards are offered:

\begin{itemize}[leftmargin=1em,itemsep=0ex]
    \item The top three teams will each receive a USD 2,500 cash prize and full travel support (covering transportation, accommodation, and registration) for one representative to present at NeurIPS.
    \item The top five teams will be invited to present their work during the competition track (15-minute slots each) and will be invited to contribute to a special issue of a leading journal (TBC).
    \item The top three teams will also be recognized as consortium authors on the subsequent NeurIPS competition publication, following the format of our previous NeurIPS competition \cite{beetl2022}.
\end{itemize}

To actively encourage newcomers and researchers from under-represented backgrounds, we will offer a dedicated prize of USD 1,000 and NeurIPS workshop registration to the best-performing first-time team led and with the majority of women or members of minority groups, adhering to NeurIPS DIA guidelines. We are committed to inclusivity and will partner with advocacy groups such as Women in AI, Black in AI, Queer in AI, and LatinX in AI to broaden our outreach. Finally, hosting a forum will foster continued engagement beyond the competition, promoting code sharing, collaboration and discussions on EEG decoding to foster collaboration and reproducible research \cite{nerve}. \vspace{-0.5em} 

\vspace{-0.5em}
\section{Resources} \vspace{-0.5em}

\paragraph{\textbf{3.1.} Organizing team}
The organization is diverse and includes scientists from different institutions and countries, including France, USA, Israel, Canada, Cuba, China, Brazil and Netherlands. The individual contributions are listed below and more details on the team can be found in \autoref{sec:biography}.
\begin{itemize}[leftmargin=1em,itemsep=0ex]
\item \textbf{Coordinators}
Bruno, Dung, Pierre, and Yahya serve as core coordinators. Isabelle, Sylvain, and Alexandre F. provide strategic oversight. Isabelle and the ChaLearn team support Codalab hosting. Amit provides access to SDSC compute resources.

\item \textbf{Data Providers: }
Michael~M. and Alexandre~F. provide the EEG datasets. Pedro and Alan: data-sharing expertise. Yahya, Scott, Oren, Amit, and Arnaud: BIDS, HED, and NEMAR expertise.

\item \textbf{Baseline Method Providers: }
Bruno and Pierre (deep learning). Aviv (machine learning for computational psychiatry). Dung and Arnaud (HBN data).

\item \textbf{Beta Testers: }
Jean-Rémi with his team his team, Yahya and Marie-Constance assist in early testing and validation of the challenge platform.

\item \textbf{Evaluators: }
Arnaud, Sylvain, Alexandre G., Isabelle Guyon and Terry Sejnowski contribute domain-specific evaluation, \emph{ensuring scientific rigor}.
\end{itemize} \vspace{-1em}

\paragraph{\textbf{3.2.} Resources provided by organizers:}

We provide HPC resources with GPUs through Mesocentre Paris-Saclay, which were used to evaluate submissions for the preliminary edition of the competition at \textsc{CodaBench.com}. These could be used for conducting during the inference phase. 

\paragraph{\textbf{3.3. Support requested:}}

We would greatly appreciate support from the NeurIPS 2025, particularly in promoting our challenge through their channels.

\appendix

\section{Biography of all team members}
\label{sec:biography}

\begin{enumerate}[leftmargin=1em,itemsep=0ex]

\item \textbf{Bruno Aristimunha}
\begin{itemize}
  \item \textbf{Affiliation}: \lisn; \\ \nerv; \ufabc
  \item \textbf{E-mail:} \href{mailto:b.aristimunha@gmail.com}{b.aristimunha@gmail.com}
  \item Research Engineer at INRIA TAU and PhD student at the University of Paris-Saclay. 
  \item Machine Learning Research Engineer specializing in deep learning and signal processing for EEG analysis. Lead Maintainer of the widely used Braindecode library and Core Developer for the Mother of all BCI Benchamark - MOABB benchmarking framework, actively shaping standards and enabling \emph{reproducible research} in brain-computer interfaces and EEG decoding. Experience in deep learning and machine learning applied to EEG decoding, validated through peer-reviewed publications, community engagement with reviewer at NeurIPS, ICLR, ICML and more journals.
  \item Core developer of Braindecode and Mother of BCI Benchmark Python toolkits. Go open-source!
\end{itemize}

\item \textbf{Dung Truong}  
\begin{itemize}
  \item \textbf{Affiliation}: Swartz Center for Computational Neuroscience, University of California, San Diego (UCSD)
  \item \textbf{E-mail:} \href{mailto:dutruong@ucsd.edu}{dutruong@ucsd.edu}
  \item Research Engineer at Swartz Center for Computational Neuroscience.
  \item Specialize in the standardization and large-scale processing of EEG data, with a research focus on developing deep learning algorithms for EEG decoding—particularly in the areas of representation learning using self-supervised methods and generative models.
  \item Core developer and maintainer of multiple open-source and neuroinformatic projects, including HED, EEGLAB, EEGDash and NEMAR.

\end{itemize}

\item \textbf{Pierre Guetschel}  
\begin{itemize}
  \item \textbf{Affiliation}: Donders Institute for Brain, Cognition and Behaviour, Radboud University, Netherlands
  \item \textbf{E-mail:} \href{mailto:pierre.guetschel@donders.ru.nl}{pierre.guetschel@donders.ru.nl}
  \item Pierre Guetschel is a PhD candidate at the Donders Institute. His research interests are on developing deep learning algorithms for EEG decoding, with a special focus on transfer learning, self-supervised learning and foundation models.
  \item Core developer of Braindecode and Mother of BCI Benchmark Python toolkits. 
\end{itemize}

\item \textbf{Seyed Yahya Shirazi}
\begin{itemize}
  \item \textbf{Affiliation}: Swartz Center for Computational Neuroscience, Institute for Neural Computation, University of California San Diego, CA, USA
  \item \textbf{E-mail:} \href{mailto:shirazi@ieee.org}{\underline{shirazi@ieee.org}}
  \item Yahya is an Assistant Project Scientist at UC San Diego, specializing in Brain-Behavior research, computational neuroscience, and neuroinformatics initiatives.
  \item Led the HBN-EEG data curation and annotation.
  \item Lead Scientist for BIDS extension proposals to EMG and Stimulus. Core member of the HED working group. Core member of the EEGLAB development team
\end{itemize}

\item \textbf{Isabelle Guyon}  
\begin{itemize}
  \item \textbf{Affiliation}: Université Paris-Saclay, France. ChaLearn, California. Google DeepMind, California.
  \item \textbf{E-mail:} \href{mailto:guyon@chalearn.org}{\underline{guyon@chalearn.org}}
  \item Isabelle Guyon is Director, Research Scientist at Google DeepMind, in detachment from her position as professor of Artificial Intelligence at Université Paris-Saclay (Orsay). She specializes in data-centric AI, statistical data analysis, pattern recognition, and machine learning. Her areas of expertise include computer vision, bioinformatics, and power systems. She has been a strong promoter of challenges and benchmarks. Her recent interests include AI-assisted human communication. Prior to joining Paris-Saclay she worked as an independent consultant and was a researcher at AT\&T Bell Laboratories, where she pioneered applications of neural networks in the 80’s with Yann LeCun and Yoshua Bengio, among others. She is president of Chalearn, a non-profit dedicated to organizing challenges in machine learning, community lead of Codalab competitions a challenge organization platform, action editor of the Journal of Machine Learning Research (JMLR), co-editor of the Data-Centric Machine Learning Research Journal (DMLR), and served as program co-chair of NIPS 2016 and general co-chair of NIPS 2017. She is the 2020 recipient of the BBVA Frontiers in Research Award together with Prof. Schoelkopf and Prof. Vapnik for contributions to kernel methods (including the invention of Support Vector Machines - SVM) and to causality in machine learning.
  \item Expert Advisor.
\end{itemize}

\item \textbf{Alan Evans} 
\begin{itemize}
  \item \textbf{Affiliation}: McGill University, Canada
  \item \textbf{E-mail:} \href{mailto:alan.evans@mcgill.ca}{\underline{alan.evans@mcgill.ca}}
  \item Distinguished James McGill Professor of Neurology and Psychiatry at McGill U.
  \item Director, EEGNet consortium; expert in structural brain network modeling. \item Co-director, Global Brain Consortium (GBC).
  \item Director, Canadian Open Neuroscience Platform
\end{itemize}

\item \textbf{Pedro Antonio Valdés-Sosa} 
\begin{itemize}
  \item \textbf{Affiliation}: Cuban Neuroscience Center (CNEURO) and the University of Electronic Science and Technology of China (UESTC)
  \item \textbf{E-mail:} \href{mailto:pedro.valdes@neuroinformatics-collaboratory.org}{\underline{pedro.valdes@neuroinformatics-collaboratory.org}}
  \item Pioneer in EEG source localization, Bayesian modeling, and brain connectivity analysis.
  \item Co-director of the Global Brain Consortium (GBC) and contributor to EEG-based BCI research.
\end{itemize}

\item \textbf{Scott Makeig} 
\begin{itemize}
  \item \textbf{Affiliation}: Swartz Center for Computational Neuroscience, University of California, San Diego (UCSD)
  \item \textbf{E-mail:} \href{mailto:smakeig@ucsd.edu}{\underline{smakeig@ucsd.edu}}
  \item Pioneer in EEG analysis and development of Independent Component Analysis (ICA) for brain signal decomposition.
  \item Leader in studying brain dynamics and mobile brain/body imaging (MoBI).
\end{itemize}

\item \textbf{Alexandre Gramfort} 
\begin{itemize}
  \item \textbf{Affiliation}: Meta Reality Labs, Paris, France
  \item \textbf{E-mail:} \href{mailto:agramfort@meta.com}{\underline{agramfort@meta.com}}
  \item Senior Research Scientist, Meta, (Ex-) Research director, HdR, Inria, MIND Team, Univ. Paris Saclay
  \item Currently senior research scientist manager at Meta Reality Labs in Paris. Works on machine learning technologies to decode surface EMG signals. Previously was research director (DR, HdR) at Inria, leading the MIND Team, known formerly as Parietal. Works on statistical machine learning, signal processing, optimization, scientific computing and software engineering with primary applications in neuroscience and biosignal processing (in particular EEG).
\end{itemize}

\item \textbf{Jean-Rémi King} 
\begin{itemize}
  \item \textbf{Affiliation}: Brain \& AI team, FAIR, Meta, Paris, France
  \item \textbf{E-mail:} \href{mailto:jeanremi@meta.com}{\underline{jeanremi@meta.com}}
  \item Research Scientist, Meta
  \item CNRS researcher at École Normale Supérieure currently detached to Meta AI, where he leads the Brain \& AI team. This team aims to identify the brain and computational bases of human intelligence, with a focus on language. For this, they develop deep learning algorithms to decode and model brain activity recorded with MEG, EEG, electrophysiology and fMRI.
\end{itemize}

\item \textbf{Terrence J Sejnowski}
\begin{itemize}
  \item \textbf{Affiliation}: Salk Institute for Biological Studies and the University of California, San Diego (UCSD)
  \item \textbf{E-mail:} \href{mailto:terry@salk.edu}{\underline{terry@salk.edu}}
  \item Co-developed the Boltzmann machine and contributed foundational work to deep learning.
  \item Instrumental in bridging neuroscience and machine learning.
\end{itemize}

\item \textbf{Marie-Constance Corsi}
\begin{itemize}
    \item \textbf{Affiliation}: NERV team, Sorbonne Université, Institut du Cerveau – Paris Brain Institute – ICM, CNRS, Inria, Inserm, AP-HP, Hôpital de la Pitié-Salpêtrière, Paris, France
    \item \textbf{Email:} \href{mailto:marie-constance.corsi@inria.fr}{\underline{marie-constance.corsi@inria.fr}}
    \item Inria research scientist at Paris Brain Institute in the NERV Lab. Her research aims to enhance Brain-Computer Interfaces (BCIs) by identifying neurophysiological markers of training and integrating multimodal data for better classifier information. Additionally, she is developing interpretable AI tools for diagnosing neurological diseases.
\end{itemize}

\item \textbf{Michael P. Milham}  
\begin{itemize}
  \item \textbf{Affiliation}: 
  \item \textbf{Email:} \href{mailto:michael.milham@nki.rfmh.org}{michael.milham@nki.rfmh.org}
  \item Michael P. Milham, MD, PhD, is an internationally recognized neuroscience researcher, a child and adolescent psychiatrist, and Director for the Center of Biomedical Imaging and Neuromodulation at the Nathan S. Kline Institute for Psychiatric Research. Dr. Milham is one of our nation’s most prolific neuroimaging researchers, with over 200 articles published since 2005 and recognition as a Clarivate Highly Cited Researcher (top .1\% for Neuroscience and Behavior) every year since 2014. He was also the recipient of the Organization of Human Brain Mapping’s highly prestigious Wiley Young Investigator Award in 2014.
\end{itemize}

\item \textbf{Alexandre R Franco}  
\begin{itemize}
  \item \textbf{Affiliation}: Data Informatics and Sharing of Knowledge Core, Child Mind Institute, USA
  \item \textbf{Affiliation}: Computational Neuroimaging Laboratories, Nathan Kline Institute for Psychiatric Research, USA
  \item \textbf{Email:} \href{mailto: alexandre.franco@childmind.org}{alexandre.franco@childmind.org}
  \item Alexandre R. Franco, PhD is the Director of the Data Informatics and Sharing of Knowledge Core at the Child Mind Institute and Director of the Computational Neuroimaging Laboratories at the Nathan Kline Institute. His research focuses on optimizing MRI and EEG data collection, data processing and data sharing.
  \item Leads the efforts for the International Neuroimaging Data-sharing Initiative.
\end{itemize}

\item \textbf{Aviv Dotan} 
\begin{itemize}
  \item \textbf{Affiliation}: Dept. of Cognitive and Brain Sciences, Ben-Gurion University of the Negev, Beer-Sheva, Israel
  \item \textbf{Email:} \href{mailto:avivdot@post.bgu.ac.il}{\underline{avivdot@post.bgu.ac.il}}
  \item Postdoctoral researcher in the Computational Psychiatry and Neurotechnology Lab at BGU.
  \item Expert in machine learning and deep learning applications to EEG data.
\end{itemize}

\item \textbf{Oren Shriki} 
\begin{itemize}
  \item \textbf{Affiliation}: Dept. of Cognitive and Brain Sciences, Ben-Gurion University of the Negev, Beer-Sheva, Israel
  \item \textbf{Email:} \href{mailto:shrikio@bgu.ac.il}{\underline{shrikio@bgu.ac.il}}
  \item Head of the Computational Psychiatry and Neurotechnology Lab at BGU.
  \item Leader in the development and application of neurotechnology.
\end{itemize}

\item \textbf{Amit Majumdar} 
\begin{itemize}
  \item \textbf{Affiliation}: San Diego Supercomputer Center (SDSC), University of California, San Diego (UCSD)
  \item \textbf{Email:} \href{mailto:majumdar@sdsc.edu}{\underline{majumdar@sdsc.edu}}
  \item Director of SDSC's Data Enabled Scientific Computing (DESC) division.
  \item Director of the Neuroscience Gateway (NSG).
\end{itemize}

\item \textbf{Sylvain Chevallier} 
\begin{itemize}
  \item \textbf{Affiliation}: A\&O/TAU team director LISN, Université Paris-Saclay, Inria TAU, France
  \item \textbf{Email:} \href{sylvain.a.chevallier@inria.fr}{\underline{sylvain.a.chevallier@inria.fr}}
  \item Full professor at the University Paris-Saclay, a board member of the DATAIA/ClusterIA Institute of Saclay, and a co-leader of the TAU team. He works on frugal learning and transfer learning, which are applied to time series analysis with applications such as BCI computing. He contributed to several open-source Python toolboxes for signal processing and machine learning using Riemannian geometry. He is leading the Codalab/Codabench framework for benchmark and data competition.
\end{itemize}

\item \textbf{Arnaud Delorme} 
\begin{itemize}
  \item \textbf{Affiliation}: University of California, San Diego (UCSD); Paul Sabatier University, France
  \item \textbf{Email:} \href{mailto:arno@ucsd.edu}{\underline{arno@ucsd.edu}}
\item Leads the EEGLAB project, a widely adopted open-source toolbox for EEG data analysis.
\item Research focuses on advanced EEG methodologies, integrating AI, and exploring the neuroscience of meditation and mind wandering.
\end{itemize}

\end{enumerate}

\section*{Acknowledgements}
This work was performed using HPC resources from GENCI--IDRIS (Grant 2025-AD011014817R1) on the Jean Zay supercomputer.

\bibliography{hbn}
\bibliographystyle{plainnat}

\end{document}